\documentclass[letterpaper,twocolumn,prl,aps,superscriptaddress,amsmath,amssymb,floatfix]{revtex4-2}
\usepackage{mathptmx}
\usepackage{color}
\usepackage{amsmath}
\usepackage{amssymb}
\usepackage{graphicx}
\usepackage{esint}
\usepackage{siunitx}
\usepackage[unicode=true,
 bookmarks=true,bookmarksnumbered=false,bookmarksopen=false,
 breaklinks=false,pdfborder={0 0 1},backref=false,colorlinks=true]
 {hyperref}
\hypersetup{
 linkcolor=magenta,urlcolor=blue,citecolor=blue,pdfstartview={FitH},hyperfootnotes=false}

\makeatletter

\usepackage{textcomp}
\usepackage{epstopdf}

\usepackage{amsfonts}

\pdfpageheight\paperheight
\pdfpagewidth\paperwidth

\usepackage{xcolor}\usepackage{soul}
\newcommand{\bra}[1]{\ensuremath{\left\langle#1\right|}}
\newcommand{\ket}[1]{\ensuremath{\left|#1\right\rangle}}

\definecolor{blue}{rgb}{0,0,1}
\definecolor{red}{rgb}{1,0,0}
\definecolor{green}{rgb}{0,1,0}

\usepackage{soul}

\makeatother

\begin{document}


\global\long\def\figurename{FIG.}%
\title{Fundamental limits of free-space microwave-to-optical frequency conversion efficiency using Rydberg atoms}
\author{Ya-Nan Lv}
\affiliation{CAS Key Laboratory of Quantum Information, University of Science and Technology of China, Hefei, Anhui 230026, P. R. China}
\affiliation{Anhui Province Key Laboratory of Quantum Network, University of Science and Technology of China, Hefei 230026, P. R. China}
\affiliation{Department of Chemical Physics, University of Science and Technology of China, Hefei 230026, China}
\author{Yan-Lei Zhang}
\email{zyl12@ustc.edu.cn}
\affiliation{CAS Key Laboratory of Quantum Information, University of Science and Technology
of China, Hefei, Anhui 230026, P. R. China}
\affiliation{Anhui Province Key Laboratory of Quantum Network, University of Science and Technology of China, Hefei 230026, P. R. China}
\affiliation{CAS Center for Excellence in Quantum Information and Quantum Physics, University of Science and Technology of China,
Hefei 230026, China}
\author{Xu-Bo Zou}
\affiliation{CAS Key Laboratory of Quantum Information, University of Science and Technology
of China, Hefei, Anhui 230026, P. R. China}
\affiliation{Anhui Province Key Laboratory of Quantum Network, University of Science and Technology of China, Hefei 230026, P. R. China}
\affiliation{CAS Center for Excellence in Quantum Information and Quantum Physics, University of Science and Technology of China,
Hefei 230026, China}
\author{Guang-Can Guo}
\affiliation{CAS Key Laboratory of Quantum Information, University of Science and Technology
of China, Hefei, Anhui 230026, P. R. China}
\affiliation{Anhui Province Key Laboratory of Quantum Network, University of Science and Technology of China, Hefei 230026, P. R. China}
\affiliation{CAS Center for Excellence in Quantum Information and Quantum Physics, University of Science and Technology of China,
Hefei 230026, China}
\author{Shui-Ming Hu}
\affiliation{Department of Chemical Physics, University of Science and Technology of China, Hefei 230026, China}
\author{Chang-Ling Zou}
\email{clzou321@ustc.edu.cn}
\affiliation{CAS Key Laboratory of Quantum Information, University of Science and Technology
of China, Hefei, Anhui 230026, P. R. China}
\affiliation{Anhui Province Key Laboratory of Quantum Network, University of Science and Technology of China, Hefei 230026, P. R. China}
\affiliation{CAS Center for Excellence in Quantum Information and Quantum Physics, University of Science and Technology of China,
Hefei 230026, China}

\date{\today}

\begin{abstract}
Efficient microwave-to-optical frequency conversion (MOC) is crucial for applications such as radiometry, electrometry, quantum microwave illumination and quantum networks. Rydberg atoms provide a unique platform for realizing free-space MOC, promising wide-bandwidth, scalable, and flexible quantum interfaces. Here, we develop a theoretical framework to evaluate the system conversion efficiency, accounting for the mismatch between microwave and optical wavelengths comparing with the atomic ensemble size. Our analysis reveals that the conversion efficiency is fundamentally limited by the focusing of the free-space microwave field, with an upper bound of about $3/16$ for diffraction-limited focusing. We propose using a microwave near-field antenna to overcome this limit. Our work provides a foundation for assessing and optimizing free-space MOC, paving the way for a variety of applications based on free-space MOC.
\end{abstract}
\maketitle


\textit{Introduction.-} Microwave and optical frequencies are two complementary wavebands that play crucial roles in modern information technologies~\cite{assoulyQuantumAdvantageMicrowave2023, bondMicrowaveImagingSpacetime2003, arumugamOpticalFiberCommunication2001}. On one hand, microwave signals are particularly useful in environment characterization, radar systems, and radiometry for detecting fundamental physics, such as axions and the cosmic microwave background radiation~\cite{bradleyMicrowaveCavitySearches2003,huCosmicMicrowaveBackground2001}. With the advent of quantum information technologies, microwaves have become essential carriers for various quantum systems, including superconducting qubits, spin qubits, and semiconductor double quantum dot qubits~\cite{burkardSemiconductorSpinQubits2023,freyDipoleCouplingDouble2012, xiangHybridQuantumCircuits2013b}. Furthermore, employing exotic microwave quantum states has the potential to realize quantum-enhanced microwave technologies with improved sensing and detection sensitivities~\cite{dingEnhancedMetrologyCritical2022a, tuApproachingStandardQuantum2023b}. On the other hand, the optical frequency domain offers advantages such as low-noise at room temperature, enabling long-distance communications and efficient single-photon counting~\cite{esmaeilzadehSinglephotonDetectorsCombining2017}. Consequently, quantum transducers that can coherently convert between microwave and optical domains are indispensable for bridging these two important frequency regimes~\cite{Han2021,laukPerspectivesQuantumTransduction2020}.

Over the past decade, significant progress has been made in developing integrated photonic waveguide or microresonator system for coherent microwave-to-optical frequency conversion (MOC). These approaches include electro-optic interaction~\cite{fanSuperconductingCavityElectrooptics2018a,Soltani2017Oct}, magneto-optic coupling~\cite{zhangOptomagnonicWhisperingGallery2016a,Hisatomi2016May}, optomechanics~\cite{Stannigel2010Nov,Bagci2014Mar,Andrews2014Apr}, and doped rare-earth atomic ensembles~\cite{OBrien2014Aug}. These platforms offer huge potential for compact, chip-integratable, and enhanced interactions due to strong light confinement, making them promising for a wide range of applications. In addition to waveguide-based systems, recent experimental demonstrations have shown remarkable advancements in MOC using cold Rydberg atoms~\cite{OBrien2014Aug,Williamson2014Nov,hattermannCouplingUltracoldAtoms2017,hanCoherentMicrowavetoOpticalConversion2018b,kumarQuantumenabledMillimetreWave2023a,vogtEfficientMicrowavetoopticalConversion2019a,tuHighefficiencyCoherentMicrowavetooptics2022c}. Rydberg atom-based schemes have emerged as promising contenders due to their strong electric dipole coupling and rich electronic structure~\cite{saffmanQuantumInformationRydberg2010c}. Additionally, the Rydberg atom system also provides unique advantages in realizing free-space MOC, which offers several advantages over waveguide-based approaches. By directly transmitting microwave signals through air, free-space conversion avoids the transmission losses associated with cables and can potentially benefit from the low cosmic background temperature~\cite{lanucaraNewObservablesCosmic2023,dhalCalculationCosmicMicrowave2023}. Moreover, Rydberg atom-based systems can work with a wide angle of free-space microwave modes, providing additional degrees of freedom for detection and communication applications, such as wide-angle receivers and phased arrays for detecting microwave signals~\cite{yanThreedimensionalLocationSystem2023}. However, achieving high end-to-end efficiency in free-space MOC remains a challenge, as most studies have focused on the internal conversion efficiency while neglecting potential geometric constraints and mismatches between the microwave and optical spatial modes~\cite{tuHighefficiencyCoherentMicrowavetooptics2022c}.

In this Letter, we aim to bridge the gap between the theoretical promise and practical realization of efficient free-space microwave-to-optical conversion using Rydberg atoms. We develop a comprehensive model that captures the fundamental limitations imposed by the mismatch between microwave and optical wavelengths, providing a realistic assessment of the achievable conversion efficiency. Through our analysis, we uncover the critical role played by microwave focusing in determining the upper bound of the conversion efficiency, which is $3/16$. We suggest to use near-field antenna to surpass the limitation. By laying the theoretical groundwork and identifying key optimization avenues, our work opens up new possibilities for harnessing the free-space microwave-to-optical conversion, ultimately enabling quantum technologies that rely on seamless integration of microwave and optical domains.

\begin{figure}    \centering{\includegraphics[width=\columnwidth]{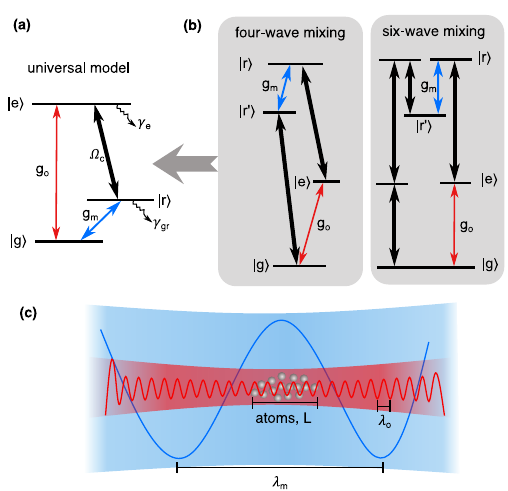}}
    \caption{Schematic of the microwave-to-optical frequency conversion (MOC) based on Rydberg atoms. (a) The $\Lambda$-type energy level diagram as the universal model for MOC. (b) The energy level diagrams of MOC based on four- and six-wave mixing processes. In (a) and (b), black arrows indicate the coherent drive fields, red arrow denotes the optical signal, and blue arrow denotes the microwave signal. (c) The illustration of the Rydberg atom ensemble coupling to the microwave and the optical fields. The blue (red) region and the corresponding solid line indicate the microwave (optical) field, with $L,\,\lambda_{\mathrm{m}},\,\lambda_{\mathrm{o}}$ remarking the ensemble length, microwave wavelength and optical wavelength, respectively.
 }
    \label{Fig1}

\end{figure}

\textit{Model.-} The simplified model for the coherent MOC in an atomic ensemble is based on a $\Lambda$-type energy level scheme, as illustrated in Fig.~\ref{Fig1}(a)~\cite{Williamson2014Nov,fernandez-gonzalvoCavityenhancedRamanHeterodyne2019a}. In this model, a strong coherent drive field (black arrow, Rabi frequency $\Omega_{\mathrm{c}}$) is applied to couple the optical excited state $\ket{e}$ and the intermediate microwave excited state $\ket{r}$. An input optical signal (red arrow) excites the  $\ket{g}\leftrightarrow\ket{e}$ transition, and the drive stimulates microwave emission (blue arrow) through the $\ket{r}\leftrightarrow\ket{g}$ transition. Conversely, a microwave input induces optical emission.

While practical implementations employ more complex energy level structures, such as the four-wave mixing~\cite{kumarQuantumenabledMillimetreWave2023a, coveyMicrowavetoopticalConversionFourwave2019b} and six-wave mixing~\cite{hanCoherentMicrowavetoOpticalConversion2018b, vogtEfficientMicrowavetoopticalConversion2019a,tuHighefficiencyCoherentMicrowavetooptics2022c,kiffnerTwowayInterconversionMillimeterwave2016a} schemes shown in Fig.~\ref{Fig1}(b), the fundamental principles remain the same. First, the transitions involved in MOC must form a closed loop, regardless how many energy levels are involved. If the transitions form a chain instead of a loop, a conversion of photon will result in a population transfer from the initial state to the end of the transition chain, and induces the entanglement between the converted photon and the excitation in the atom ensemble which means decoherence of the converted fields when tracing out the MOC device. Second, the essential mechanism of MOC is nonlinear frequency mixing, where coherent beam-splitter-type interaction between the optical and microwave field are stimulated by external coherent drives. Specifically, for an $m$-wave mixing process, a total of $m-2$ coherent drives are employed to stimulated the MOC process. In waveguide-based MOC schemes, three-wave mixing is realized by electro-optics and magneto-optics effects\cite{fanSuperconductingCavityElectrooptics2018a,Soltani2017Oct,zhangOptomagnonicWhisperingGallery2016a}, and four-wave mixing is realized in electro-opto-mechanical systems~\cite{midoloNanooptoelectromechanicalSystems2018}. These nonlinear susceptibility of dielectrics are usually broadband but weak, the waveguide and resonator structures are utilized to enhance the interactions. In contrast, atomic ensemble system exploits near-resonant transitions to achieve strong nonlinear optical susceptibility for multiple wave mixing. Based on the two principles, all atom-based MOC schemes can be reduced to the effective model shown in Fig.~\ref{Fig1}(a). 

Figure~\ref{Fig1}(c) illustrates the cold atom ensemble mediating the coherent coupling between the free-space microwave and optical beams. This system is particularly interesting due to the difference in the wavelengths of microwave ($\lambda_{\mathrm{m}}\sim$mm) and optical ($\lambda_{\mathrm{o}}\sim\mathrm{\mu m}$) photon compared to the size of the atom ensemble ($L\sim$cm), with the relationship $\lambda_{\mathrm{m}}\gg L \gg \lambda_{\mathrm{o}}$. This disparity in scales leads to distinct collective light-matter interaction behavior, prompting a separate investigation of the microwave beam-atom and optics beam-atom couplings in the following.

\begin{figure}
    \centering{\includegraphics[width=\columnwidth]{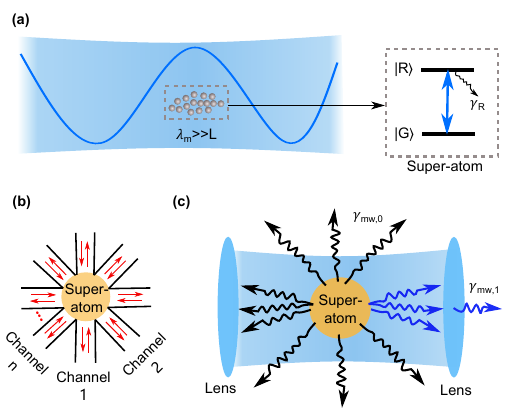}}
    \caption{(a) Schematic of the coupling between Rydberg atom transition with microwave field, with the atom ensemble treated as a super-atom. (b) The multiple channel model illustrates the coupling of free-space continuum modes with the super-atom. (c) The free-space microwave field modes coupled to the super-atom can be divided into two parts, one can be collected (blue arrows) and the other can't (black arrows). }
    \label{Fig2}
\end{figure}

\textit{Microwave coupling.-} As depicted in Fig.~\ref{Fig1}(a), the atoms couple to the microwave field through Rydberg energy levels $\ket{r'}$ and $\ket{r}$. Since  $L\ll\lambda_{\mathrm{m}}$, atoms can be approximated as a point and the collective behavior of atoms could be treated as a single harmonic oscillator with a weak input microwave field, referred to as a ``super-atom" in Fig.~\ref{Fig2}(a). The excitation in $N$-atom ensemble can be represented by bosonic operator $S^\dagger=\frac{1}{\sqrt{N}}\sum_j\sigma_j^{+}$ according to the Holstein-Primakoff approximation~\cite{kuruczMultilevelHolsteinPrimakoffApproximation2010b}, with subscript $j$ denote $j$-th atom and $\sigma^{+}=\ket{r}\bra{g}$ is the raising operator. Here, the approximation applies when $N\gg1$ and the excitation number is small, and also the Rydberg blockade effect between atoms is negligible due to the large mean distance between excited atoms compared to the blockade radius~\cite{urbanObservationRydbergBlockade2009, gaetanObservationCollectiveExcitation2009}. The ground state of the system is denoted as $\ket{G}=\ket{gg...gg}=\ket{g}_{\otimes N}$, the single excitation state is given by $\ket{R}=S^\dagger \ket{G}$, with a collective amplitude decay rate due to spontaneous emission (SE) is $\gamma_{\mathrm{R}}=N\gamma_{gr}$~~\cite{grimesDirectSingleshotObservation2017,haoObservationBlackbodyRadiation2021}, where $\gamma_{gr}$ is the SE rate of a single atom transition from $\ket{r}$ to $\ket{g}$.

The conversion process in Rydberg atoms can be described by the scattering between different radiation channels of the super-atom, as described in Fig.~\ref{Fig2}(b), each channel represents a continuum of modes, such as a Gaussian beam in free space. The conversion efficiency is determined by the ratio of  the SE rate into a target continuum mode to that of  the undesired modes~\cite{solomonsUniversalApproachQuantum2024}.
The multi-channel coupling is be described by the Hamiltonian
\begin{equation}     H_{m}=\omega_{s}S^{\dagger}S+\sum_{p=1}^{n}\sum_{k}[\omega_{p,k}b_{p,k}^{\dagger}b_{p,k}+g_{k}b_{p,k}^{\dagger}S+g_{k}^{*}b_{p,k}S^{\dagger}],\nonumber
\end{equation}
where $\omega_{s}$ is the effective transition frequency of the super-atom, $\omega_{p,k}$ and $b_{p,k}$ are the frequency and bosonic operator of the microwave mode with wave-vector $k$ in the $p$-th channel, respectively, given the coupling strength $g_{p}$ with the super-atom. Applying the Markovian and Born approximations, the steady-state solution (see Supplementary Material~\cite{sm} for details) for the super-atom operator under external fields $b_{p,\mathrm{in}}$ with frequency $\omega$ is given by
\begin{equation}
    S(\omega)=\frac{-i\sum_{p=1}^{n}\sqrt{2\gamma_{p}}b_{p,\mathrm{in}}}{-i(\omega-\omega_{s})+\gamma_{\mathrm{R}}'+\sum_{p=1}^{n}\gamma_{p}},
    \label{eq:ss_S}
\end{equation}
where  $\gamma_{p}$ represents the SE rate of super-atoms into $p$-th channel, and $\gamma_{\mathrm{R}}^{\prime}$ is the amplitude decay rate due to non-radiative decay and also radiative decay into other frequencies. For a Gaussian beam with a beam waist $w_0$, the one-dimensional continuum gives $\gamma_{\mathrm{mw,1}}=(3\lambda_{\mathrm{m}}^2/4\pi^2 w_0^2)\gamma_{\mathrm{R}}$. The sum of the decay rate into all channels gives to the collective SE rate, i.e., $\sum_{p=1}^{n} \gamma_{p} =\gamma_{\mathrm{R}}$. In free space, all microwave field modes can be divided into two parts, as shown in Fig.~\ref{Fig2}(c), one part is the first channel ($p=1$) corresponds to the beam that can be excited and collected (blue arrows), while the remaining modes compose the other part (black arrows). According to the input-output theory, the reflectivity and the transmission of the microwave signal are given by $r_{\mathrm{mw}}(\omega)=\sqrt{2\gamma_{\mathrm{mw,1}}} S(\omega)/b_{1,in}$ $t_{\mathrm{mw}}(\omega)=1-i\sqrt{2\gamma_{\mathrm{mw,1}}} S(\omega)/b_{1,in}$.

\begin{figure} \centering{\includegraphics[width=\columnwidth]{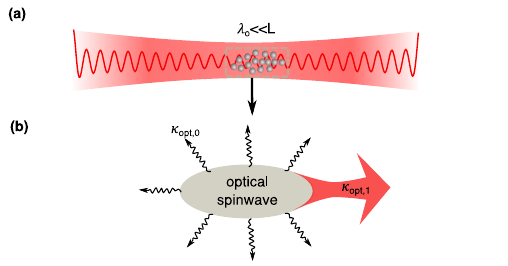}}
    \caption{(a) Schematic of the collective coupling between atom ensemble and the optical field, with the optical wavelength being much smaller than the ensemble length ($\lambda_{\mathrm{o}}\ll L$). (b) The collective decay of spin-wave excitation in the ensemble leads to directional emission into a free-space optical beam, with a spontaneous emission rate of $\kappa_{\mathrm{opt,1}}$.
    }
    \label{Fig3}
\end{figure}

\textit{Optical coupling.-} Since the ensemble size is much larger than the optical wavelength $L\gg\lambda_{o}$, the collective effect of the optical coupling differs from the microwave coupling, as shown in Fig.~\ref{Fig3}(a). To describe the collective interactions, we introduce the concept of a collective spin-wave mode, defined as $E_{\vec{k}}=\frac{1}{\sqrt{N}}\sum_{j}e^{i\vec{k}\cdot\vec{r}_{j}}\sigma_{ge,j}$, where $\sigma_{ge,j}$ is the localized atomic jump operator at position  $\vec{r_j}$ ,  $\vec{k}$ is the wave-vector. The spin-wave mode is essentially a Fourier transformation of localized atomic excitations. Consequently, the coupling between atoms and the optical modes reads
\begin{equation}
    H_{o}=\frac{\lvert g_{o}\rvert}{\sqrt{N}}\sum_{\vec{k}_{0}}\sum_{\vec{k}_{1}}\sum_{j=1}^{N}e^{i(\vec{k}_{0}-\vec{k}_{1})\cdot\vec{r}_{j}}e^{i(\omega_{a,\vec{k}_{0}}-\omega_{ge})t}a_{\vec{k}_{0}}^{\dagger}E_{\vec{k}_{1}}+h.c.,
    \label{eq:5_H_3D_2}
\end{equation}
where $g_{o}$ is the coupling strength between single atom and the plane wave optical modes, $a_{\vec{k}_{0}}$ notes the annihilation operators of optical photons, and $\vec{k}_{0,1}$ denotes the wave-vector of the optical (spin-wave) mode. This Hamiltonian shows that there are multiple collective atomic excitation couples with various optical modes, making the general treatment of the optical coupling with the atomic ensemble complex. Instead of using a multiple channel model as studied with the microwave coupling case [Fig.~\ref{Fig2}(b)], we introduce a free-space optical radiation beam that matches the target spin-wave mode, as illustrated in Fig.~\ref{Fig3}(b). The essential difference for the coupling model is originates from the distinct scale of the wavelength and the atomic ensemble, where the constructive interference between individual atoms leads to directional optical radiation, while the dipole-dipole coupling between atom is negligible. The free-space radiation beam has a size significantly larger than the wavelength, and the optical setup could be efficiently modulated for perfectly matching the profile of radiation mode. Then, by approximating the spatial density function of the atoms ensemble as Gaussian distribution, we can reconstruct the one-dimensional free-space beam mode that emitted by the spin-wave excitation, with its bosonic operator as $r_{k_{0}}^{\dagger}=\frac{1}{A}\sum_{k_{0,z}}^{k_{0}}\sum_{\theta}e^{-\frac{1}{2}(k_{0}^{2}-k_{0,z}^{2})\sigma^{2}}a_{\vec{k}_{0}}^{\dagger}$. Here, $A$ is a normalization factor, $\sigma$ is the width of the atom cloud in optical light propagation ($z$) direction, and more details are provided in Ref.~\cite{sm}. For this radiation mode, we can derive the coupling between the spin-wave mode with the continuum as the SE rate $\kappa_{\mathrm{opt}}=\frac{\overline{\mathrm{OD}}}{4}\gamma_{e}$, where $\gamma_{e}$ is the SE rate of a single atom, and  $\overline{OD}$ is the mean optical depth of the atomic ensemble in $z$-direction. In this work, we consider an $\overline{\mathrm{OD}}\gg1$~\cite{sm, heAtomicSpinwaveControl2020, jiSuperradiantDetectionMicroscopic2023}. Compared with the multiple-channel model for microwave coupling, the optical coupling can be described by a single-channel model. However, other effects, such as the diffusion of atoms that changes the locations and collisions between atoms, can induce an extra decay rate $\kappa_{\mathrm{opt,0}}$ to the spin-wave.

\begin{figure}
    \centering{\includegraphics[width=\columnwidth]{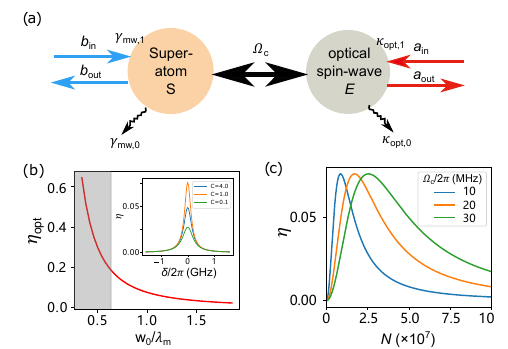}}
    \caption{(a) Schematic of the conversion between microwave and optical signals by coupling the super-atom and spin-wave modes. (b) The fundamental limit on optimal conversion efficiency ($\eta_{\mathrm{opt}}$) imposed by the microwave beam waist ($w_0$), Inset: the dependence of $\eta$ on the frequency detuning $\delta$ for different cooperativities ($C$). (c) Conversion efficiency ($\eta$) as a function of atom number $N$ for various Rabi frequencies $\Omega_c$.}
    \label{Fig4}
\end{figure}

\textit{Conversion efficiency.-} As illustrated in Fig.~\ref{Fig4}(a), by combining the input-output models for the atom ensemble's coupling to microwave and optical free-space beams,  we can realize MOC by introducing a coherent coupling between the super-atom and spin-wave excitations. Through the control light field with Rabi frequency $\Omega_{c}$, the MOC system can be described by the Hamiltonian
\begin{align}
    H=&\delta E^{\dagger}E+\Omega_{c}(S^{\dagger}E+SE^\dagger)\nonumber\\
    &+[\sqrt{2\gamma_{\mathrm{mw,1}}}S^\dagger b_{\mathrm{in}}+\sqrt{2\kappa_{\mathrm{opt,1}}}E^\dagger a_{\mathrm{in}}+h.c.],
    \label{eq:H_conversion_1}
\end{align}
where $\delta=\omega_{e}-\omega_{a}$ is the detuning between the optical transition ($\omega_e$) and the optical signal frequency $\omega_a$. We assume the microwave signal is on-resonance with the microwave transition and the external drive matches the frequency difference between optical and microwave signals. The details about the derivation of the above conversion Hamiltonian are provided in the Supplementary Material~\cite{sm}.

The MOC model enables bidirectional frequency conversion between microwave and optical photons. For either input, a part of the excitation can be converted to the desired electromagnetic wave frequency band with a quantum efficiency given by
\begin{equation}
    \eta=\frac{\gamma_{\mathrm{mw,1}}}{\gamma_{\mathrm{R}}^{\prime}+\gamma_{\mathrm{R}}}\frac{\kappa_{\mathrm{opt,1}}}{\kappa_{\mathrm{opt,0}}+\kappa_{\mathrm{opt,1}}}\frac{4C}{(1+C)^{2}+\delta^{2}/\kappa_{\mathrm{opt}}^{2}},
    \label{eq:eta}
\end{equation}
where $C=\Omega_{c}^{2}/(\gamma_{\mathrm{R}}'+\gamma_{\mathrm{R}})(\kappa_{\mathrm{opt,0}}+\kappa_{\mathrm{opt,1}})$ is the cooperativity for the coherent coupling between the super-atom and the spin-wave modes. It is worth noting that this expression resembles the results for other cavity-enhanced MOC~\cite{Han2021}. The first two terms represent extraction factors, determined by the ratio between the external coupling rate and the total decay rate of the modes, corresponding to the impedance matching or mode matching in practical experiments. The third term is the internal conversion efficiency, reaching unity when $C=1$ and $\delta=0$. Therefore, the maximum achievable $\eta$ is intrinsically limited by the extraction factors.

Considering that the internal decay rate $\kappa_{\mathrm{opt,0}}$ can be much smaller than the emission rate of the optical spin-wave, i.e., $\kappa_{\mathrm{opt},0}\ll \kappa_{\mathrm{opt},1}$, in cold atom ensemble by optimizing its temperature and density, and the ratio between Rydberg state's intrinsic decay and stimulated decay ($\gamma_{\mathrm{R}}'/\gamma_{\mathrm{R}}$) can optimized to approach zero via increasing the coherent drive strengths, the eventual $\eta$ is limited by the extract ratio of the microwave port as
\begin{equation}
    \eta\leq \frac{\gamma_{\mathrm{mw,1}}}{\gamma_{\mathrm{R}}}=\frac{3\lambda_{\mathrm{m}}^2}{4\pi^2w_0^2}.
    \label{fundamentallimit}
\end{equation}
According to the diffraction limit of a free-space Gaussian beam ($w_0>2\lambda_{\mathrm{m}}/\pi$), the upper bound of $\eta$ is $3/16$. Figure~\ref{Fig4}(b) shows the achievable optimal $\eta_{\mathrm{opt}}$ for a given microwave beam waist $w_0$ in free-space, with the shaded area denoting the region forbidden by the diffraction limit. To mitigate this fundamental limit of free-space MOC, one approach is to introduce a microwave cavity to enhance the collection of microwave signals, at the cost of sacrificing the wide-angle and broadband response of atoms. Alternatively, an elongated atom ensemble can be employed to increase the microwave interaction length and go beyond the $L\ll \lambda_{\mathrm{m}}$ limitation. Nevertheless, this requires a few cm-long atom ensemble with a considerable cold atoms density (see Supplementary Materials for more details~\cite{sm}), imposing significant challenges in practical experiments. We suggest using near-field antennas to concentrate the free-space microwave mode as a promising solution. This approach could break the $3/16$ efficiency limit while preserving the potential advantages of free-space MOC by leveraging the sub-diffraction confinement of microwave fields near the antenna surface, achieving stronger coupling between the microwave photons and the atomic ensemble.

To evaluate the MOC performance under different conditions, we numerically simulated a $^{87}$Rb atom ensemble~\cite{tuHighefficiencyCoherentMicrowavetooptics2022c}. The microwave beam, with a beam waist  $w_0=\lambda_{\mathrm{m}}$,  is coupled to the Rydberg transition $\lvert 40 P_{3/2}\rangle\to\lvert 39 D_{5/2}\rangle$, while the optical beam is coupled to the transition $\lvert 5 S_{1/2}\rangle\to\lvert 5 P_{3/2}\rangle$. According to the fundamental limit [Eq.~(\ref{fundamentallimit})], we can only achieve an efficiency of $7.6\%$. The inset of Fig.~\ref{Fig4}(b) shows the conversion efficiency against the detuning for various $C$, showing the conversion bandwidth of about $0.32,0.26,0.45\,\mathrm{GHz}$ for $C=4,1,0.1$ respectively, which is mainly determined by $\kappa_{\mathrm{opt},1}$ with $\Omega_{c}/2\pi=20$~MHz. For a fixed atomic transition and $\delta=0$, $C$ linearly increases with $\Omega_{c}^{2}N$, with the corresponding OD of the atom ensemble estimated with a transverse atomic cloud waist of $66\,\mathrm{\mu m}$. Figure~\ref{Fig4}(c) shows that $\eta$ reaches its maximal value for an optimal $N$. A larger $N$ means a larger $C>1$, inhibiting MOC due to coupling-induced coherent oscillations when exchanging the excitation between super-atom and spin-wave, equivalent to a spectral splitting in the response making the system's optimal working condition deviating from $\delta=0$. Similarly, $C$ is also determined by $\Omega_c^2$, and the optimal $N$ increases with increasing $\Omega_c$.

\textit{Conclusion.-} Our theoretical framework provides the upper bound of the achievable free-space MOC efficiency, which arise from the mismatch between the microwave wavelength and the size of atomic ensemble. Our work suggests that near-field antennas can be helpful in realizing high-efficient quantum interfaces based on free-space atom ensembles, which have the potential to enable a wide range of applications in quantum technologies, including quantum communication, quantum sensing, and hybrid quantum systems~\cite{gisinQuantumCommunication2007,degenQuantumSensing2017,xiangHybridQuantumCircuits2013b}. Furthermore, the theoretical methods developed here can be extended to other applications involving disparate length scales of electromagnetic modes, such as the fundamental limitation of the detection sensitivity of electrometry~\cite{dingEnhancedMetrologyCritical2022a, tuApproachingStandardQuantum2023b}, and imaging and spectroscopy in the terahertz regime\cite{jepsenTerahertzSpectroscopyImaging2011,wadeRealtimeNearfieldTerahertz2017}.

\nocite{Wall2008}
\nocite{zhuLightScatteringDense2016}
\nocite{brionAdiabaticEliminationLambda2007a}
\nocite{paulischAdiabaticEliminationHierarchy2014}
\nocite{Wang2019}

\begin{acknowledgments}
We thank K.-Y. Liao, H. Yan, and Z.-T. Liang for discussion. This work was funded by the National Key R\&D Program (Grant No.~2021YFA1402004), and the National Natural Science Foundation of China (Grants (No.~12347129, U21A20433, 92265108, 92265210, 12393822 and 11874342).  This work was also supported by the Fundamental Research Funds for the Central Universities and USTC Research Funds of the Double First-Class Initiative. Y.-N.L. was also supported by Postdoctoral Fellowship Program of CPSF. The numerical calculations in this paper were performed on the supercomputing system in the Supercomputing Center of the University of Science and Technology of China. This work was partially carried out at the USTC Center for Micro and Nanoscale Research and Fabrication.
\end{acknowledgments}


%

\end{document}